\documentclass[twocolumn]{autart}
 \newcommand{\comments}[1]{}
\usepackage{subfigure}
\usepackage{graphicx}
\usepackage{epsfig}
\usepackage{mathptmx}
\usepackage{times}
\usepackage[cmex10]{amsmath}
\usepackage{amssymb}
\usepackage{dsfont}
\usepackage{mathrsfs}
\usepackage{cases}
\usepackage{extpfeil}
\usepackage{txfonts}
\usepackage{mathrsfs}
\usepackage{stmaryrd}
\usepackage{epstopdf}
\usepackage{amsfonts}
\usepackage{cite}
\usepackage{color}
\usepackage{stfloats}
\usepackage{float}
\usepackage{algorithmic}
\usepackage{algorithm}
\newtheorem{theorem}{Theorem}

\newtheorem{remark}{Remark}

\newtheorem{assu}{Assumption}
\newtheorem{defi}{Definition}

\belowdisplayskip \abovedisplayskip%
\begin{document}
\newcommand{\nn}{\nonumber}
\newcommand{\bea}{\begin{eqnarray}}
\newcommand{\eea}{\end{eqnarray}}
\newcommand{\beas}{\begin{eqnarray*}}
\newcommand{\eeas}{\end{eqnarray*}}

\newcommand{\tcr}{\textcolor{red}}
\newcommand{\tcb}{\textcolor{blue}}
\newcommand{\tcm}{\textcolor{magenta}}
\newcommand{\tcg}{\textcolor{green}}
\newcommand{\tcc}{\textcolor{cyan}}

\begin{frontmatter}
\title{A hierarchical control framework for autonomous decision-making systems: Integrating HMDP and MPC}
\thanks[footnoteinfo]{This work was supported by the UK Engineering and Physical Sciences Research Council (EPSRC) Established Career Fellowship (EP/T005734/1).\\}
\author[LAB]{Xue-Fang Wang}, \ead{xw259@leicester.ac.uk.}
\author[CAB]{Jingjing Jiang}, \ead{j.jiang2@lboro.ac.uk}
\author[CAB]{Wen-Hua Chen$^*$} \ead{w.chen@lboro.ac.uk}
\thanks[footnoteinfo]{$^*$ Corresponding author: Wen-Hua Chen.}
\address[LAB]{School of Engineering, University of Leicester, UK}
\address[CAB]{Department of Aeronautical and Automotive Engineering, Loughborough University, UK}

\begin{keyword}
Hybrid Markov decision process (HMDP), autonomous decision-making systems, unified hierarchical control framework, model predictive control (MPC), safety and optimality
\end{keyword}

\begin{abstract}
This paper proposes a comprehensive hierarchical control framework for autonomous decision-making arising in robotics and autonomous systems. In a typical hierarchical control architecture, high-level decision making is often characterised by discrete state and decision/control sets. However, a rational decision is usually affected by not only the discrete states of the autonomous system, but also the underlying continuous dynamics even the evolution of its operational environment. This paper proposes a holistic and comprehensive design process  and framework for this type of challenging problems, from new modelling and design problem formulation to control design and stability analysis.  It addresses the intricate interplay between traditional continuous systems dynamics utilized at the low levels for control design and discrete Markov decision processes (MDP) for facilitating high-level decision making. We model the decision making system in complex environments as a hybrid system consisting of a controlled MDP and autonomous (i.e. uncontrolled) continuous dynamics. 
Consequently, the new formulation is called as hybrid Markov decision process (HMDP).
The design problem is formulated with a focus on ensuring both safety and optimality while taking into account the influence of both the discrete and continuous state variables of different levels.
With the help of the model predictive control (MPC) concept, a decision maker design scheme is proposed for the proposed hybrid decision making model. By carefully designing key ingredients involved in this scheme, 
it is shown that the recursive feasibility and
stability of the proposed autonomous decision making scheme are guaranteed. 
The proposed framework is applied to develop an autonomous lane changing system for intelligent vehicles. Simulation shows it exhibits a promising ability to handle diverse behaviors in dynamic and complex environments.  It is envisaged that the proposed framework is applicable to autonomous decision making for a wide range of robotics and autonomous systems where both safety and optimality are key considerations. 
\end{abstract}
\end{frontmatter}

\section{Introduction}
\subsection{Background}
The emergence of autonomous systems necessitates a novel shift in control methodologies due to its increasing complexity, capabilities and wide applications (see, for example,  \cite{antsaklis2020autonomy,sharma2022retail,chen2022perspective,seetohul2023augmented,xu2018reinforcement,li2023aid,bertram2022fast,yan2023surviving}). 
Unlike traditional control systems that predominantly focus on the designing and implementing  control algorithms for continuous system dynamics, autonomous systems require a holistic approach that includes  high-level decision-making within an admissible decision set. 
This necessitates the exploration and development of novel control methodologies that can effectively govern these autonomous decision-making systems.

It is now established that the achievement of extensive decision-making capabilities is a fundamental element to accomplish full automation of vehicles.
\begin{figure}[htb!]
	\centering
	\includegraphics[width=6cm]{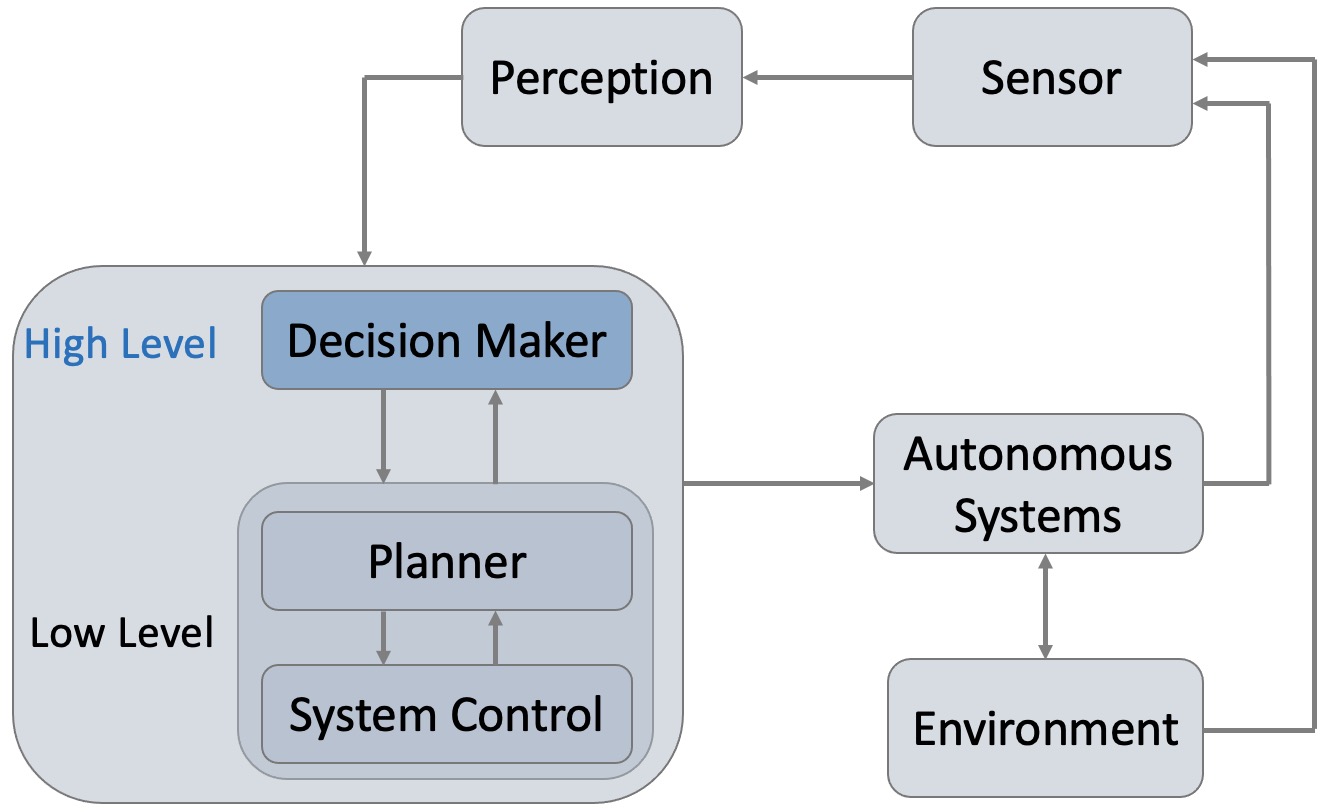}
	\caption{Hierarchical autonomous control architecture that may consist of several levels.}
	\label{autonomousarchi}
\end{figure}
In the realm of autonomous driving (as illustrated in Fig. \ref{autonomousarchi}), with decision maker, we refer to a module that, taking as input a representation of the environment around the vehicle, selects a  high-level  maneuver from a pre-defined decision set and guides the vehicle with  appropriate driving behavior. 
The decision-making module communicates its choice to the \emph{planner}, which calculates a spatial trajectory and speed profile by
leveraging the environment representation and the current vehicle state. 
The path planning module is designed to meet the vehicle's dynamics, enabling the execution of the selected maneuver.
Thus, decision-making in autonomous driving concerns with abstract and intentional actions, while delegate
the physical actuation of the vehicle (\emph{e.g.}, steering angle, acceleration) to the low-level modules (\emph{i.e., System Control} in Fig.~\ref{autonomousarchi}).

\subsection{Motivation}
\begin{figure}[htb!]
	\centering
	\includegraphics[width=5cm]{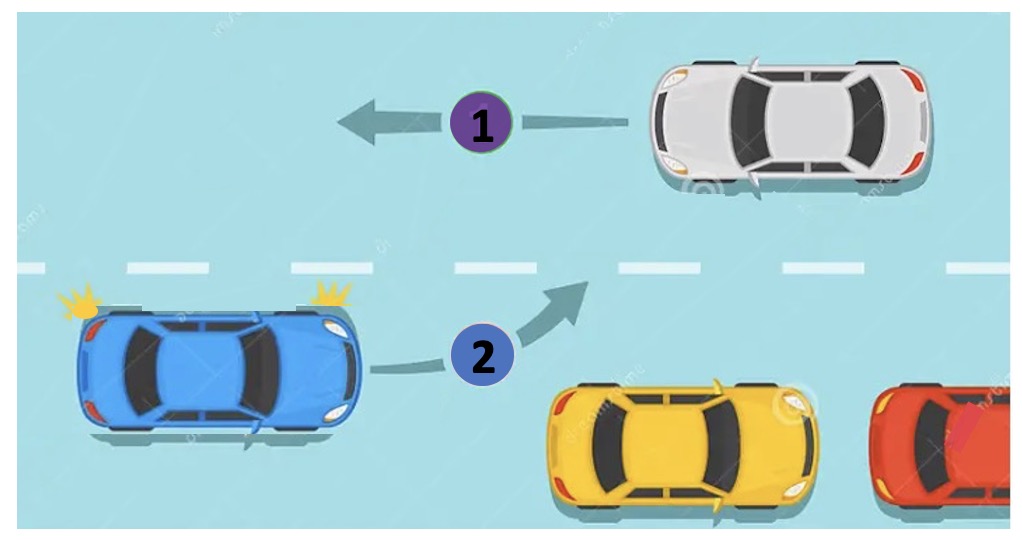}
	\caption{Wait behind the parked cars to give way to oncoming vehicles (Blue one is the ego vehicle).}
	\label{autonomouswait}
\end{figure}

\begin{figure}[htb!]
	\centering
	\includegraphics[width=5cm]{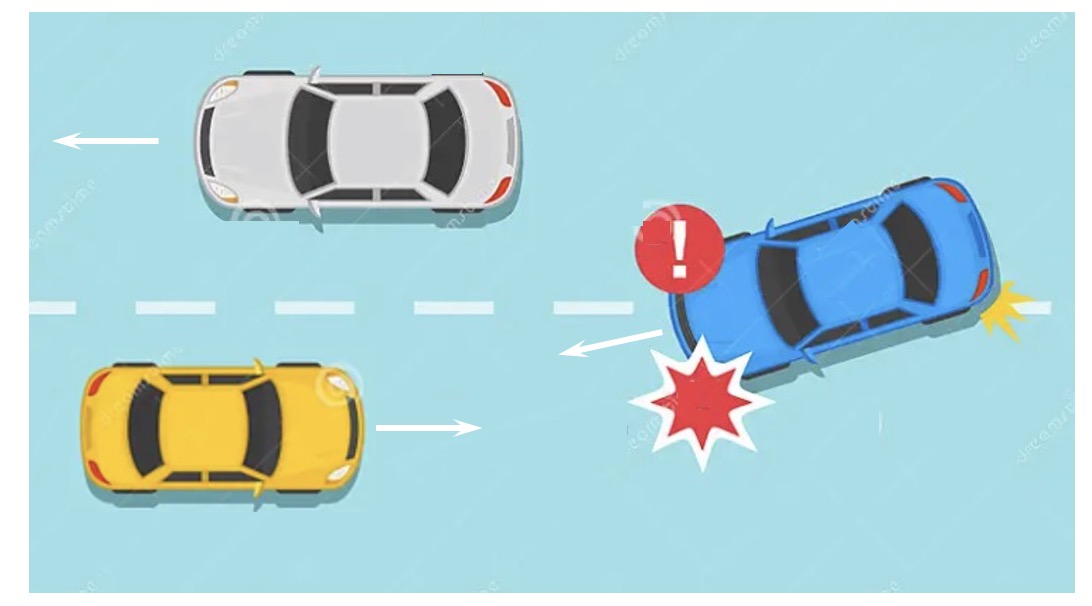}
	\caption{Overtaking car collision with oncoming car (Blue one is the ego vehicle).}
	\label{autonomousovertaking1}
\end{figure}

\begin{figure}[htb!]
	\centering
	\includegraphics[width=6cm]{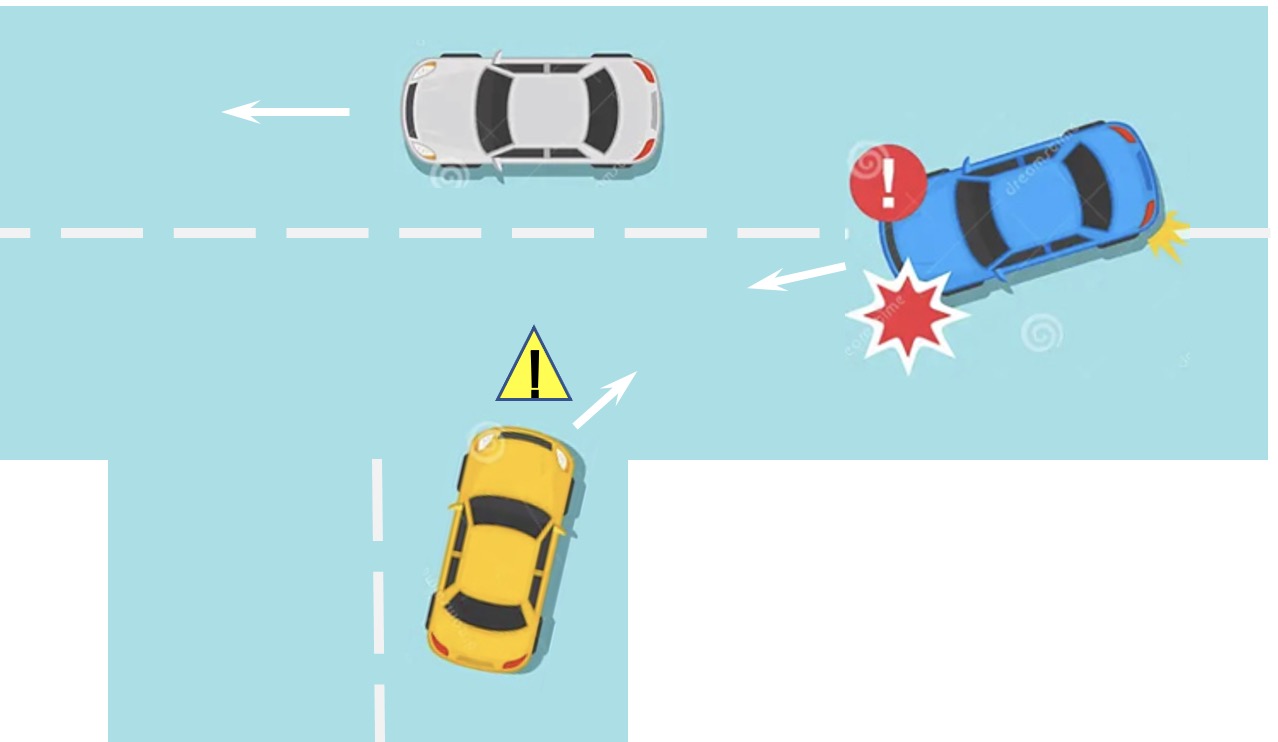}
	\caption{Abandon overtaking to give way to the sudden-emerging vehicle (Blue one is the ego vehicle).}
	\label{autonomousovertaking2}
\end{figure}

Examples of autonomous overtaking with physical interaction to different environments are depicted in Fig.~\ref{autonomouswait} - Fig. \ref{autonomousovertaking2}.
In these autonomous overtaking scenarios on two-lane country roads in the presence of oncoming vehicles, 
the ego vehicle has lower priority on the opposite lane. This means that the ego vehicle must give way to oncoming vehicles (see, Fig. \ref{autonomouswait}). 
In addition, if the distance from the front car is too small, the field of view of the ego vehicle is partially blocked and crash will happen on the road (see, Fig. \ref{autonomousovertaking1}). 
However, if the distance from the front car is too large, the overtaking time will be quite long, resulting in more uncertainties during the overtaking maneuver.
Overall, in such scenarios, a typical set of maneuvers that the decision maker can select from is \{
``follow the current path’’, ``stop at a certain location’’, ``change lane to right or left’’, and ``give way to oncoming vehicles’’\}.
Clearly, the design of high-level decision maker is crucial to generate an  ``appropriate maneuver’’ for various situations, and it is responsible for handling  situations and events that are discrete in nature.
Because of this intrinsic duality between the high-level discrete event and the 
low-level continuous dynamics  of the vehicle, such interplay phenomenon is called \emph{hybrid}. 
For such \emph{hybrid} interplay, the methods designed for pure switched systems are unsuitable (see, for example, \cite{wang2016stability,wu2022policy}). 
This implies that it makes sense to consider decision-making dynamics that
are modeled by hybrid dynamical systems where vehicles with continuous states have the capability of synthesizing discrete event-based decision-making strategies.

Such scenarios are not limited to autonomous driving. The hierarchical autonomous control architecture as shown in Fig. \ref{autonomousarchi} is quite generic, for example, see \cite{verbakel2020decision,you2019advanced}.
The decision-making process in autonomous systems involves complex tasks such as perception, cognition, planning, and execution, which go beyond the scope of traditional control systems.  
Consequently, while traditional control systems remain integral to the functioning of autonomous systems, they are no longer dominant. 
Instead, they act as subordinate elements, providing supportive functionalities to facilitate the realization of autonomous behavior. 
By embracing this broader perspective and considering the intricate interplay between traditional control systems and high-level decision-making processes, we can unlock the transformative capabilities of autonomous systems, leading to unprecedented levels of efficiency, safety, and autonomy in various real-world applications. 

Indeed, the applied scenarios are complex that they are either impossible or inappropriate to describe decision-making processes with conventional system models such as differential or difference equations \cite{zhang2022integrated}.
Consequently, the integration of traditional control systems with high-level decision-making processes poses significant challenges. 
This convergence involves combining two distinct domains of automation and decision-making, each with its own complexities and requirements.
Hence, an integrated system must address challenges related to security and reliability, ensuring correctness and reliability of the decision-making and control processes.
However, to the best knowledge of authors, there is no  unified framework for autonomous decision-making systems by considering the intricate interplay between traditional control systems and higher-level decision-making processes. 
Therefore, it is significant to design a unified control framework that overcomes these challenges. 

\subsection{Related work}
The existing studies on decision making can be broadly categorized into three main groups, \emph{i.e.,} rule-based methods \cite{jula2000collision,hwang2022autonomous,zhang2022integrated}, learning-based approaches \cite{hou2015situation}, and model-based techniques \cite{bucsoniu2018reinforcement,zhou2017hierarchical}. 
For instance, paper \cite{jula2000collision} proposed a rule-based minimum safety spacing calculation method in lane changing and lane merging scenarios based on vehicle kinematics. This method has gained widespread adoption in various applications, such as risk assessment in trajectory planning \cite{luo2016dynamic}, maneuver modeling \cite{butakov2014personalized}, and overtaking \cite{ngai2011multiple}. 
However, it is important to note that the method's robustness is compromised when dealing with different lane changing trajectories, as the calculation of the critical collision point varies across trajectory formulations.
In contrast, the learning-based method \cite{hou2015situation} has demonstrated the ability to generate human-like behaviors.
Nevertheless, its efficacy in critical driving scenarios can be significantly limited by the availability of training data, which is often insufficient.
In other words, rule-based and learning-based methods each have their strengths and limitations, making them suited for particular use cases but not universally generalizable solutions.

Markov decision process (MDP) offers a promising alternative tool to develop decision making approaches for autonomous systems  (see, for example, \cite{verbakel2020decision,yun2022cooperative,feng2016synthesis,beard2022safety,likmeta2020combining,bucsoniu2018reinforcement,zhou2017hierarchical}).
 In general, the previous works along the MDP approach represent a significant advancement to enhance the likelihood of safe actions but fall short of providing absolute guarantees due to the persistent risk of unsafe actions occurring (see, for example, \cite{bucsoniu2018reinforcement,zhou2017hierarchical}).
Following the contributions of work \cite{larsen2016online,ashok2019sos}, paper \cite{goorden2023guaranteed} applied the mathematical modeling framework of hybrid Markov decision process (HMDP) to propose a method for synthesizing safe controllers for continuous-time sampled switched systems, where the analytical solution for the state trajectories is available. This framework was applied to cruise control of autonomous driving and management of storm water detention ponds and obtained stable results under the assumption that the environment switches periodically.
Hence, taking various environmental factors into consideration, it is a vital task to integrate safety constraints into the decision-making process to ensure the autonomous system operates safely. 
This challenge becomes particularly pronounced when the safety constraints involve both the states of the underlying dynamic systems at a low level and the states of the decision-making model at a high level. 
In other words, to optimise the performance while ensuring safety, both the \emph{continuous} underlying dynamics and the \emph{discrete} dynamics describing high-level behaviour have to be taken into account simultaneously in making timely and rational decision in response to uncertainty and change of the operational environment.

\subsection{Contributions}
To fill the gap, and motivated by the aforementioned observations and recognizing the abundance of excellent works at low-level path planning and dynamics control \cite{ji2023tripfield,ji2016path,rosolia2022mixed,schwarting2018planning,goorden2023guaranteed}, this paper specifically focuses on decision making at a high level.
We first formulate the decision-making problem for autonomous systems as a control problem based on MDP.
Subsequently, we design a unified hierarchical control framework, enhancing an MDP with an underlying autonomous dynamic system.
This system encapsulates the essential continuous dynamic behavior of the autonomous system, contingent upon the discrete state determined by the MDP.

Real-time decision making is formulated as an optimisation problem for a new hybrid system consisting of both continuous and discrete states. 
A Model Predictive Control (MPC) like scheme is proposed, where at each step, the optimal decision is generated by solving an optimisation problem based on the real-time continuous and discrete states and the information of the operational environment. 
This framework offers an optimal and safe solution for decision making in complex environments.
By leveraging the predictive capabilities of MPC (referring to \cite{chen1998quasi,michalska1993robust,chen2003optimal,du2022observer}), the proposed control framework enables the system  of anticipating future states and adjusting control actions accordingly.
By utilizing MDP models, autonomous systems can account for the highly changing and uncertain environments in their decision-making processes
due to the ability of MDP to model uncertainty, handle dynamic situations, deal with exploration-exploitation trade-offs, define optimality under uncertainties, and leverage the Markov property. 
Compared with finite state machine with a predetermined set of rules (see, for example, \cite{xing2021hazard,palatti2021planning,hwang2022autonomous,jaswanth2022autonomous}), these properties make MDPs a powerful and flexible framework for decision-making in complex and unpredictable scenarios.
This integration significantly enhances the framework's ability to handle dynamic and uncertain environments, enabling safer and more effective decision making.
The main contributions of this paper can be summarized as follows.
\begin{enumerate}
\item A unified modelling approach is presented to articulate a hierarchical control framework in autonomous decision-making systems.
This approach captures the intricate interplay between dynamic systems at a low level and MDP models at a high level.
This means that the decision-making process incorporates not only the MDP states used at the high-level decision making but also takes into account the states of the underlying dynamic systems at a low level. 
In addition, inspired by \cite{goorden2023guaranteed} and to distinguish it from simple classic MDP, piecewise linear systems or switched systems, we call our new framework as Hybrid MDP (HMDP).

\item Inspired by MPC, a new autonomous control solution for high level decision making is proposed by carefully specifying the cost function and constraints (e.g. safety) in terms of both continuous and discrete states.

\item The recursive feasibility and stability of the proposed framework can be guaranteed under mild conditions without requirements of terminal constraints.

\item The safety and optimality of the proposed autonomous decision-making system is guaranteed when it interacts with dynamically changing environments. Moreover, to demonstrate the effectiveness of the proposed control framework, an application scenario with simulations are investigated. 

\end{enumerate}

\subsection{Organisation}
The remainder of this paper is structured as follows. Formulation of high-level decision making with HMDP for autonomous decision-making systems is introduced in Section \ref{formulation}. In Section \ref{solution}, an MPC-based algorithm  used to solve HMDP control problem is proposed. In Section \ref{stability}, the recursive feasibility and stability of the proposed control algorithm are discussed. In Section \ref{simulations}, the proposed optimal solution is evaluated by simulations on autonomous lane changing. 
Conclusions are presented in Section \ref{conclusions}.

\section{Formulation of high-level decision making with HMDP}
\label{formulation}
In the field of conventional control systems, the central objective revolves around devising controllers that effectively meet predetermined performance metrics of the system at hand. 
However, when deal with autonomous systems at the high-level decision-making layer, the conventional control framework used at the low level serves merely as  auxiliary support to the overarching cognitive processes that govern autonomous behavior. 
For instance, when autonomous systems engage at high-level decision-making, the selection of actions necessitates careful consideration of the physical model described by traditional control systems. 
This integration ensures the paramount safety of autonomous systems in real-world applications. 
Further elaboration on this aspect will be provided in the subsequent subsections.

\subsection{System and environment dynamics in hierarchical control framework}

In this paper, we explore a hierarchical framework comprising high-level decision making and low-level control, as shown in Fig. \ref{autonomousarchi}. 
We only focus on high-level decision making, thus we assume that at the low level, the controller is well-designed and can perfectly implement the commands from high level. 
More specifically, consider an underlying continuous dynamic system at the low level as
\begin{align}\label{originaldynamics}
x(k+1)=g(x(k),u(k)), ~x(k)\in\mathcal{X}.
\end{align}
At each operational mode $s(k)$, an ideal well-designed low level controller $u(k)=\Gamma_{s(k)} (x(k),\Xi(k))$ is designed and implemented, where $\Xi(k)\in\mathcal{E}$ denotes environment state and it is described as follows:
\begin{align}\label{environment}
\Xi(k+1)=g_{e}(\Xi(k),x(k)), ~\Xi(k)\in\mathcal{E}.
\end{align}
At a high level,  we can abstract the behaviours of the dynamic system under the designed low level controller $u(k)=\Gamma_{s(k)} (x(k),\Xi(k))$ as 
\begin{align}\label{simplifieddynamics}
x(k+1)=\tilde{g}_{s(k)}(x(k),\Xi(k)), ~x(k)\in\mathcal{X},\ \Xi(k)\in\mathcal{E},
\end{align}
where $s(k)\in \mathcal{S}$ denotes the MDP state which represents the operational modes at the high level, satisfying
\begin{align}\label{HMDPdynamics}
 s(k+1)=f(s(k),\pi(s(k),x(k),\Xi(k))),
\end{align}
where  $\pi(\cdot,\cdot,\cdot)$ means that the choice of the high-level policy which depends on not only the current state $s(k)$ of MDP but also the current continuous state $x(k)$ of system (\ref{simplifieddynamics}) and environment state $\Xi(k)$. 

It follows from (\ref{simplifieddynamics}) and (\ref{HMDPdynamics}) that we model the decision making system in complex environments as a hybrid system consisting of a controlled MDP and autonomous (\emph{i.e.,} uncontrolled, but continuous) dynamics. 
We apply the mathematical framework of HMDP 
to model such a hybrid system. 
The detailed modelling with HMDP is given in the following subsection.

\subsection{Modelling of high-level decision making with HMDP} 

The high-level decision-making mechanisms of autonomous systems are critical in enabling intelligent responses to dynamic and uncertain environments. 
These systems can utilize advanced algorithms to make informed decisions and execute actions that optimize performance based on predefined objectives. 
However, in such dynamic and uncertain environments, a significant challenge lies in abstracting the behaviors of both the autonomous system and its surroundings. 
This abstraction is necessary to facilitate decision-making without relying on complex dynamic models 
because a rational decision is usually affected by not only the discrete states of the autonomous system, but also the underlying continuous dynamics even the evolution of its operational environment.
In tackling this challenge, one approach that holds great promise is the integration of HMDP into the decision-making framework of autonomous systems. 

HMDP provides a powerful modelling tool, not only describing traditional systems whose dynamics are represented by continuous- or discrete-time state space models, but also modelling discrete event phenomena \cite{bellman1957markovian,sutton1998introduction,goorden2023guaranteed}. 
By incorporating MDP models, autonomous systems can account for the highly changing and uncertain environments in their decision-making processes since MDP allows the autonomous system to gather information about the environment while simultaneously exploiting the knowledge gained to optimize its decision. 

A finite-horizon HMDP $\mathcal{M}$ is a tuple $(\mathcal{S},\mathcal{X},\mathcal{A}, f,J)$, 
where $\mathcal{S}\in\mathbb{R}^{|\mathcal{S}|}$ is the set of MDP states, 
$\mathcal{X}\in\mathbb{R}^{|\mathcal{X}|}$ is the set of autonomous system states,
$\mathcal{A}\in\mathbb{R}^{|\mathcal{A}|}$ is the set of actions, 
$f:\mathcal{S}\times\mathcal{A}\rightarrow\mathcal{S}$ is the state transition function defining next state given current state and action, 
$J(s,a):\mathcal{S}\times\mathcal{A}\rightarrow\mathbb{R}$ is the scalar cost for taking action $a$ in state $s$.
Note that, as explained in Section 2.1, the selection of action $a$ not only depends on the MDP state $s$, but also on the autonomous system state $x$.
In this paper, we assume that all the elements of $\mathcal{M}$ are known and the state transitions are \emph{Markovian}, 
meaning that the state at time step $k$ depends only on the state and actions at time step $k-1$.
To be more specific, the elements of $\mathcal{M}$ are given as follows:
\begin{itemize}
	\item \emph{State Space $\mathcal{S}$.} 

	\item \emph{Action Space $\mathcal{A}$ and Policy $\pi$.} 


In our settings with the deterministic situation, policy $\pi$ is mapping the discrete MDP state $s(k)\in\mathcal{S}$, continuous state $x(k)\in\mathcal{X}$ and the environment state $\Xi(k)\in\mathcal{E}$ to a control action $a(k)\in\mathcal{A}$,
\emph{i.e.}, $\pi:\mathcal{S}\times \mathcal{X} \times \mathcal{E}\rightarrow\mathcal{A}$.

 \item  \emph{State Transition Function.} 
We use $S_i\xrightarrow[]{\text{$a(k)$}}S_{j}$ with $S_i,S_{j}\in \mathcal{S},a(k)\in \mathcal{A}$ to denote state transition from  $S_i$ to $S_{j}$ under action $a(k)$.
According to the states transition function, we can re-write the relationship using the following function:
	\begin{align}\label{MDPstatetransitationmodel}
	    s(k+1)=f(s(k),a(k)), ~k=0,1,...,
	\end{align}
where $s(k)\in\mathcal{S}$ and $a(k)\in \mathcal{A}$ are state and control action at time step $k$. 
The action $a(k)$ is the function of $(s(k),x(k),\Xi(k))$.

\item  \emph{Set of constrained states $\mathcal{S}_{f,x}$}. $\mathcal{S}_{f,x
}\subset\mathcal{S}$ is defined as a set of discrete states satisfying all safety and other constraints. These constraints may be related to both continuous state $x$ from (\ref{simplifieddynamics}) and discrete state $s$ from (\ref{HMDPdynamics}).  An action $a(k)$ must be chosen from a nonempty action set such that  $s(k+1) \in \mathcal{S}_{f,x}$. 

\item 
The cost of applying $a(k)$ at state $s(k)$ is denoted by $J(s(k),a(k))$, and is assumed to be non-negative:
\begin{align}\label{costfunctionassu}
0\leq J(s(k),a(k))\leq \infty,~~s(k)\in\mathcal{S}, ~a(k)\in\mathcal{A}.
\end{align}
In this paper, when the task is accomplished at a certain state, we call such  state as the \emph{goal state}.
The cost $J(s(k),a(k))$ equals to zero \emph{if and only if} the {goal state} is achieved.

The cost  of a policy $\pi$ which drives the state from the initial value $s_0$ to the \emph{goal state}, denoted by $\bar{J}(s_0)$, is given as $\bar{J}(s_0)=\sum^{\infty}_{k=0}J(s(k),\pi(s(k),x(k),\Xi(k)))$.

\end{itemize}

\begin{figure}[htb!]
	\centering
	\includegraphics[width=8.5cm]{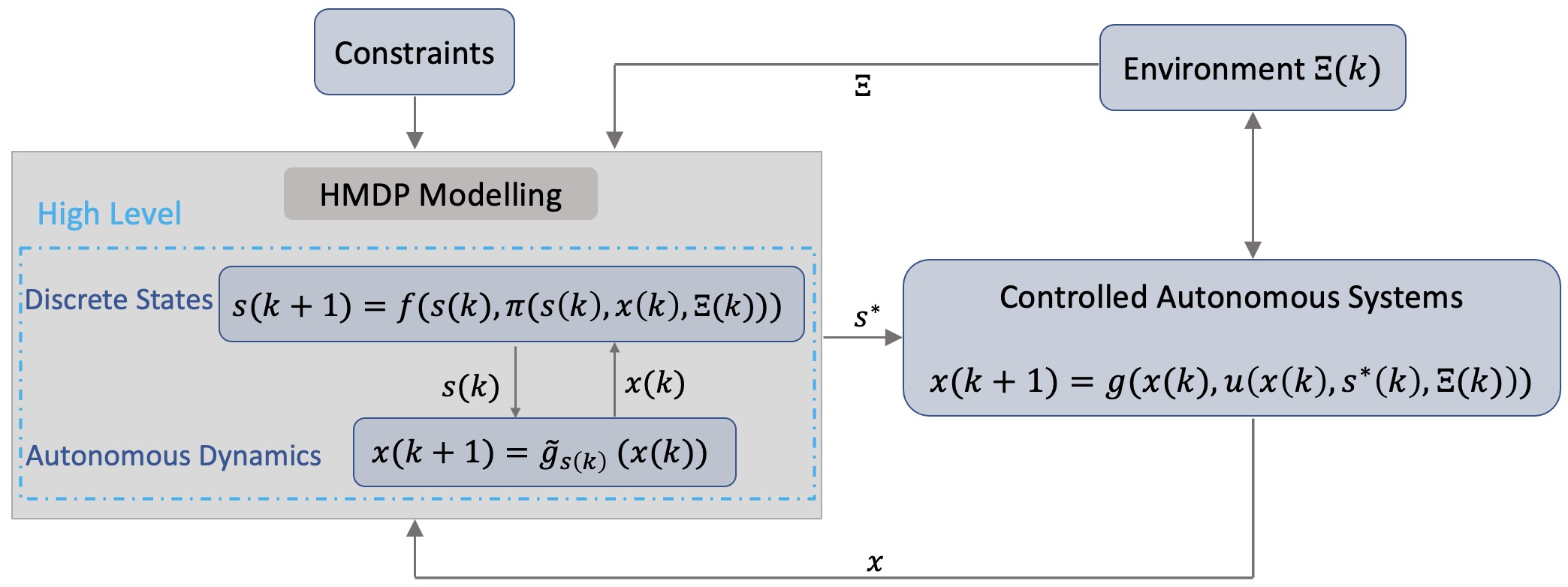}
	\caption{Hybrid decision-making systems where  $s^*$ is the high-level optimal operational mode transmitted to the low level.}
	\label{modellingstructure}
\end{figure}

The above descriptions about the high-level decision making system can be illustrated by Fig. \ref{modellingstructure}. 
From Fig. \ref{modellingstructure}, it can be clearly seen that the proposed hybrid decision making model consists of a controlled MDP and autonomous continuous dynamics.
It addresses the intricate interplay between traditional system dynamics utilized at the low levels for control design and discrete MDP for facilitating high-level decision making. 
Note that ``Autonomous Dynamics" in Fig. \ref{modellingstructure} is a simplified representation of the low level ``Controlled Autonomous systems''.

The proposed unified hybrid decision-making framework holds significant potential across various applied scenarios, encompassing high-level strategic decision-making and resolution of low-level technical issues. The hierarchical autonomous control architecture arises in a wide range of applications, including autonomous driving scenarios \cite{zhang2022integrated,hwang2022autonomous}, unmanned aerial vehicles (UAVs) for surveillance \cite{yun2022cooperative}, economic problems \cite{chizhov2011markov}, and manufacturing robots \cite{castane2023assistant}. Therefore, our proposed HMDP formulation provides a generic modelling framework to facilitate the design and analysis of high-level decision making involved in these autonomous systems.


\subsection{Formulating high-level decision making as an optimal problem}

The utilization of the HMDP framework facilitates the estimation of expected costs associated with various actions, empowering the system to make decisions that minimize the cumulative cost over time. 
To accomplish the assigned tasks successfully and ensure safety, it becomes crucial to abstract the behaviors of both the autonomous system and its surrounding environment. 
Moreover, it is imperative to carefully incorporate all relevant information, including safety constraints, into the decision-making process.
Therefore, by considering the complicated interplay between physical systems used in the low level control and high-level decision-making processes, the high-level decision making can be formulated as the following optimisation problem:
\begin{align}\label{MDPproblem}
&\min_{a(\cdot)}\sum^{\infty}_{k=0}J(s(k),a(k)),\nn\\
s.t. ~&s(k+1)=f(s(k),\pi(s(k),x(k),\Xi(k))),\nn\\
&x(k+1)=\tilde{g}_{s(k)}(x(k)), ~x(k)\in\mathcal{X},\nn\\
&\Xi(k+1)=g_{e}(\Xi(k),x(k)), ~\Xi(k)\in\mathcal{E},\nn\\
&s(k+1)\in \mathcal{S}_{f,x},\nn\\
&a(k)\in \mathcal{A}.
\end{align}
\begin{remark}
The formulation of the optimisation problem in Eq.(\ref{MDPproblem}) implies that the high-level decision making problem has a dual focus on ensuring both safety and optimality while taking into account the influence of both the discrete state variables and physical system states across different hierarchical levels. 
This formulation (\ref{MDPproblem}) presents an intricate interplay between traditional system dynamics utilized at the low levels for control design and discrete states in  MDP to facilitate high-level decision making. 
Consequently, the decision making system in complex environments is conceptualized as a hybrid system, comprising a controlled MDP (\emph{i.e.,} discrete states) and dynamic systems (\emph{i.e.,} discrete-time or continuous-time systems). 
This is different from switched systems \cite{wu2022policy}, where the switching signals can undergo abrupt changes while its states can only evolve continuously.
In our HMDP model (\ref{MDPproblem}), not only does the switching signal or action exhibit jumping, but the states within the HMDP can also jump. 
This is attributed to the expanded state dimensionality of the HMDP, which encompasses not only MDP states but also includes states from the underlying dynamic system. 
In other words, the HMDP not only incorporates MDP states (\ref{HMDPdynamics}) but also is extended to include states from the lower-level dynamic system (\ref{simplifieddynamics}) and states from the operation environment (\ref{environment}).
\end{remark}

\begin{assu}
\label{baselinepolicy}
Suppose that there exists a baseline policy sequence  that can drive the state from any initial state $s_0$ to the goal state while satisfying the safety and other constraints such as described in (\ref{MDPproblem}).
\end{assu}

\begin{remark}
Assumption \ref{baselinepolicy} is satisfied for many practical systems. 
This paper focuses on a problem that is deemed solvable; hence, it is reasonable to assume the existence of a solution. In many engineering problems such as autonomous driving, the rule-based methods have been widely used (e.g. \cite{zhang2022integrated}) for decision making at a high level for autonomous systems. Other approaches such as the Euler method in conjunction with \emph{UPPAAL TIGA} \cite{goorden2023guaranteed} also exist. These methods can be incorporated into our framework as a means to providing a baseline policy.

\end{remark}

\section{MPC-based solution to HMDP problem}
\label{solution}
The optimal decision making problem formulated in Eq.(\ref{MDPproblem}) suffers two shortcomings. First 
 autonomous systems must make safe and optimal decisions when operating in a highly dynamic and uncertain environment. It is important to keep update both the status of the environment and of the autonomous systems and adjust the decision accordingly. Secondly, it is intractable to solve an optimisation problem with an infinity horizon in real-time. To address these two issues, we resort to the receding horizon approach in MPC where then optimisation problem with an infinity horizon is truncated to that of an finite horizon, and the associated optimisation problem is repeatedly solved after regularly update the state and the environment information.
 
An MPC-based controller will be designed for the hybrid decision-making system HMDP described in Fig. \ref{modellingstructure}.
By leveraging the predictive capabilities of MPC, the system can anticipate future states and optimize control actions accordingly. 
This integration enhances the framework's ability to handle dynamic and uncertain environments.To generate safe and optimal decisions, a control algorithm based on MPC is proposed as Algorithm 1 in this section.


\begin{algorithm}\label{alg}
 \begin{tabular}{p{7.75cm}}
\textbf{ /*HMDP-based High-level decision making*/}
 \begin{enumerate}
     \item Initialisation: At initial time step, \emph{i.e.}, $k=0$, initialize the states 
$s_0$ and $x_0$. 
Given prediction horizon $N_{h}$ and sampling time $T_h$ for high level.
\item {Find the optimal action $a^{*}(i;k)$ that minimize the cost function 
\begin{align}\label{optimalcostfunc}
 \sum^{N_h-1}_{i=0}J(s(i;k),a(i;k))+\bar{J}(s_{N_h}),   
\end{align}
subject to the constraints
\begin{align}\label{optimalconstraints}
 &s(i+1;k)=f(s(i;k),\pi(s(i;k),x(i;k),\Xi(i;k))),\nn\\
&x(i+1;k)=\tilde{g}_{s(i;k)}(x(i;k)), ~x(i;k)\in\mathcal{X},\nn\\
&\Xi(i+1;k)=g_{e}(\Xi(i;k),x(i;k)), ~\Xi(i;k)\in\mathcal{E},\nn\\
&s(i+1;k)\in\mathcal{S}_{f,x},\nn\\
&a(i;k)\in \mathcal{A}, ~i=0,1,...N_h-1.
\end{align}
}
\item Execute the first action $a^{*}(0;k)$.
\item At low level, check the optimal command $s^{*}$ transmitted from high level.


 \item Output the ego vehicle's physical information $x$ obtained at low level to high level.

 \item After time duration $T_h$, $k\leftarrow k+1$, go to Step (2).
 \end{enumerate}
 \end{tabular}
\caption{\textbf{ MPC-based Solution to the Optimal HMDP Problem}} 
\end{algorithm}

\begin{remark}
In algorithm 1, $\bar{J}(s_{N_h})$ is the terminal cost that covers the cost-to-go. According to Assumption \ref{baselinepolicy} we know that there exists a baseline policy $\bar{\pi}$ that can be used to calculate $\bar{J}(s_{N_h})$ which is given as follows:
\begin{align}\label{indicatorfunction}
\bar{J}(s_{N_h})=\sum^{\infty}_{k={N_h}}J(s(k),\bar{\pi}(s(k),x(k),\Xi(k))).
\end{align}
Based on Assumption \ref{baselinepolicy} and the fact that the cost equals to zero when the \emph{goal state} is achieved, we know that $\bar{J}(s_{N_h})< \infty$.


\end{remark}


\section{Properties of HMDP: Recursive feasibility and stability}
\label{stability}

\begin{defi}\label{stabledefi}
The modeled HMDP system consisting of (\ref{HMDPdynamics}) and (\ref{simplifieddynamics}) is said to be stable under a designed policy $\pi(\cdot,\cdot,\cdot)$ if the \underline{goal state} is achieved eventually.
\end{defi}

First, let us demonstrate that Algorithm 1 consistently yields a feasible control action throughout its operation, provided that it starts with a feasible control action.

\begin{theorem}\label{recursiveana}
Suppose that Assumption \ref{baselinepolicy} hold. If Algorithm 1 is feasible at time step $k=0$, then it is feasible at all time steps $k\in\{1,2,\cdots\}$.
\end{theorem}

\textbf{Proof.}
The optimal nominal policy and the associated predicted states at time step $k$ are denoted by:
\begin{align}\label{optimalk}
&a^{*}(k):=(a^{*}(0;k),a^{*}(1;k),\cdots, a^{*}(N_h-1;k));\nn\\
&s^{*}(k):=(s^{*}(1;k),s^{*}(2;k),\cdots,s^{*}(N_h-1;k),s^{*}(N_h;k)).
\end{align}
Since $s^{*}(N_h;k)\in \mathcal{S}_{f,x}$, according to Assumption \ref{baselinepolicy} we know that there exists an action $a^{*}(N_h;k)=\pi^{*}(s^{*}(N_h;k),x^{*}(N_h;k),\Xi^{*}(N_h;k))$
such that the state $s^{*}(N_h+1;k)\in \mathcal{S}_{f,x}$ at time step $k$.

Correspondingly, we construct the following sequences which are feasible at time step $k+1$
\begin{align}\label{optimalk+1}
&a(k+1):=(a^{*}(1;k),\cdots,a^{*}(N_h-1;k),a^{*}(N_h;k));\nn\\
&s(k+1):=(s^{*}(2;k),\cdots,s^{*}(N_h-1;k),s^{*}(N_h;k),s^{*}(N_h+1;k)).
\end{align}
The proof is completed. \hfill $\Box$

In general, achieving optimality may not guarantee the stability of the closed-loop Markov Chain. 
Ensuring the stability of the algorithm becomes crucial when utilizing HMDP to model the autonomous operating process. 
This is vital to guarantee the reliability and accuracy of the model output, thereby enhancing the practicality of the proposed solution. 
Consequently, for the successful implementation of autonomous operations, the development and adoption of a stable HMDP algorithm are imperative to effectively control the autonomous operating behavior.

\begin{theorem}
Suppose that Assumption \ref{baselinepolicy} hold. Then Algorithm 1 provides a solution which stabilizes the HMDP.
\end{theorem}

\textbf{Proof.} 
According to the conclusion of Theorem \ref{recursiveana}, we are ready to compare the optimal value function $V^{*}_{N}$ at time step $k$ with the cost $V_{N}$ under the feasible sequences at time step $k+1$.

From (\ref{optimalcostfunc}), the optimal value function at time step $k$ is:
\begin{align*}
V^{*}_{N}(k)&=\sum^{N-1}_{i=0}J(s^*(i;k),a^{*}(i;k))+\bar{J}(s^*_N)\nn\\
&=J(s^*(0;k),a^{*}(0;k))\nn\\
&+\sum^{N-1}_{i=1}J(s^*(i;k),a^{*}(i;k))+\bar{J}(s^*_N).
\end{align*}
The cost function under the feasible sequences at time step $k+1$ is given by:
\begin{align*}
V_{N}(k+1)&=\sum^{N-1}_{i=1}J(s^*(i;k),a^{*}(i;k))+\hat{J}(s_N).
\end{align*}

According to (\ref{indicatorfunction}), (\ref{optimalk}) and (\ref{optimalk+1}), we can conclude that 
\begin{align*}
\hat{J}(s_N)\leq \bar{J}(s^*_N).
\end{align*}

Therefore, it follows that 
\begin{align*}
V^{*}_{N}(k+1)-V^{*}_{N}(k)&\leq V_{N}(k+1)-V^{*}_{N}(k)\nn\\
& \leq -J(s^*(0;k),a^{*}(0;k)),
\end{align*}
which implies that $V^{*}_{N}(k+1)-V^{*}_{N}(k)\leq 0$. 
In addition, $V^{*}_{N}(k+1)-V^{*}_{N}(k)=0$ if and only if $J(s^*(0;k),a^{*}(0;k))=0$, \emph{i.e.,} $s^*(0;k)$ is located at the \emph{goal state}.
Therefore, according to Definition \ref{stabledefi},  the conclusion follows. \hfill $\Box$

\section{Application to Autonomous Lane Changing}
\label{simulations}

The effectiveness of the proposed framework is evaluated through an application to Autonomous Lane Changing with simulations. 
\begin{figure}[htb!]
	\centering
	\includegraphics[width=5cm]{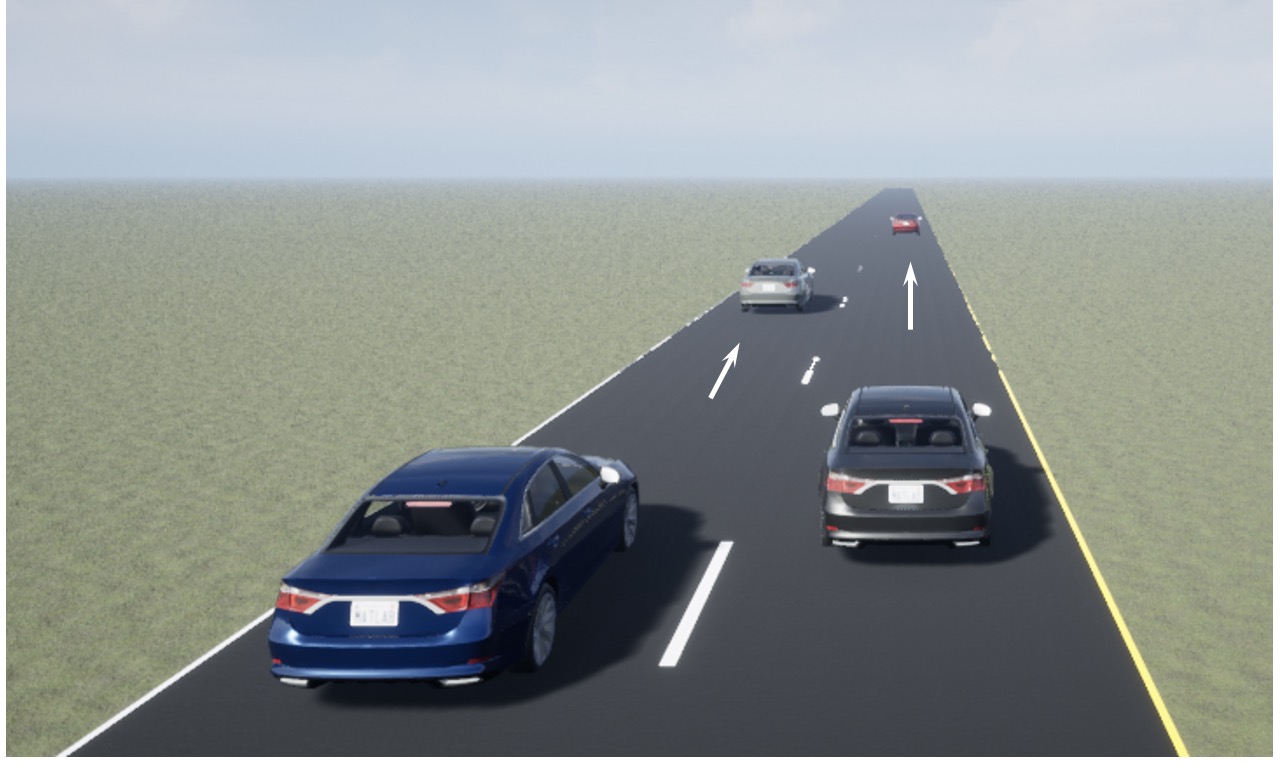}
	\caption{Testing environment designed by Driving Scenario Designer.}
	\label{scenario}
\end{figure}
To compare with the rule-based method proposed in \cite{zhang2022integrated}, we consider a similar driving scenario: There is a stationary vehicle (\emph{i.e.,} red one) ahead of the ego vehicle (\emph{i.e.,} black one) on the right lane, and there are two moving vehicles (\emph{i.e.,} blue and grey ones) on the left lane (see Fig. \ref{scenario}).
The ego vehicle needs to take a quick lane changing maneuver to avoid potential  collision. 
For the driving scenario, we use Driving Scenario Designer provided by MATLAB to set the testing environment. 
Both lanes are set to be 3.8 meters wide and all the vehicles are 2.5 meters wide.
Moreover, according to the rules given in \cite{zhang2022integrated}, it follows that Assumption  
\ref{baselinepolicy} is validated and the stable baseline policy exists.

To illustrate the effectiveness of our high-level decision making, we utilize a kinematic model in the path-planning layer, which is an MPC-based path planner,  according to the decisions made in the high level. 
This implies that the high-level decision-making and path-planning layers will be implemented at different frequencies.

The following widely used kinematic model is considered in the implementation at the low level path planning
\begin{align}\label{kinematic}
\dot{x}&=v\cos \theta\nn\\
\dot{y}&=v\sin \theta\nn\\
\dot{\theta}&=\frac{v \tan\delta}{l}\nn\\
\dot{v}&=a,
\end{align}
where $z=[x,y,\theta,v]^T$ are the states of the model and $u=[\delta, a]^T$ are the control inputs. 
Specifically, $(x,y)$ are the global coordinates of the centre point of the rear axle, $\theta$ is the heading angle of the vehicle body with respect to the $x$ axis, $v$ is the linear forward velocity, $\delta$ is the steering angle of the front wheel with respect to the vehicle's longitudinal line and $a$ is the longitudinal acceleration. 
$l$ is the distance between the front axle and the rear axle.

\subsection{HMDP modelling}
In this section, we will discuss the HMDP modelling approach to abstracting the behaviors of both the ego vehicle and the surrounding vehicles to facilitate decision-making without resorting to their complicated dynamics models.
We consider the ego vehicle as a decision maker, who interacts with the dynamic environment. 
It receives a representation of the environment's state and the state of the ego vehicle at each time step, and exerts a safe action on the environment that may change its future state.
The goal of the ego vehicle is to minimize the cost over future actions. 
To execute a quick lane changing maneuver safely and successfully, the ego vehicle needs to navigate through a sequence of discrete event states. 
These event states represent key decision points and actions that the vehicle must go through to successfully complete the maneuver.

\begin{figure}[htb!]
	\centering
	\includegraphics[width=4cm]{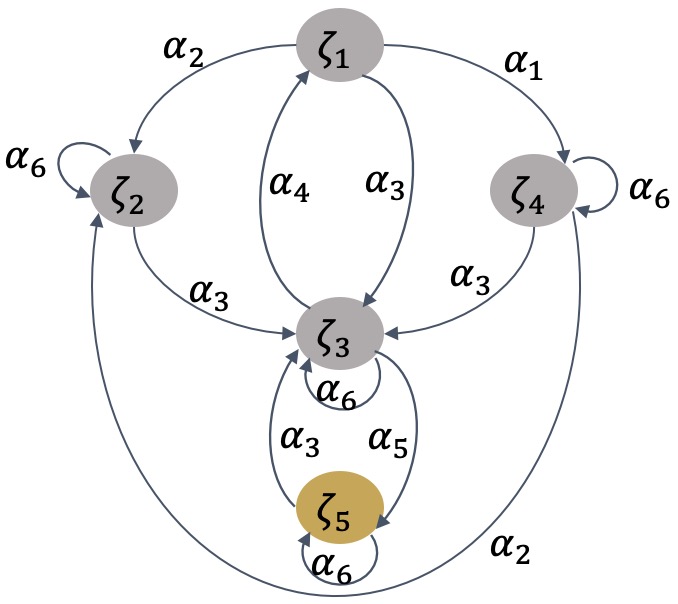}
	\caption{MDP-based state transitions}
	\label{MDPtransation}
\end{figure}

Taking inspiration from the logical diagram presented in \cite{zhang2022integrated} for decision making with four states (namely, Initialization, Braking, Quick lane change, and Acceleration), we also contemplate adopting these same states as MDP states. 
To enhance the safety of ego vehicles navigating through dynamic environments, it is crucial to emphasize that we are introducing an additional MDP state called \emph{Return}.
This state allows the ego vehicle to revert to its original lane in cases where the adjacent lane proves unsuitable for lane change continuation after a lane change attempt has been initiated. 
The MDP states transition is given in Fig. \ref{MDPtransation}.
\begin{itemize}
    \item State Space: $\mathcal{S}=\{Cruise~(\zeta_1), Braking~(\zeta_2), \\
Quick ~lane~ change~(\zeta_3), Acceleration~(\zeta_4), Return~(\zeta_5)\}$;

\item Action Space: $\mathcal{A}=\{Speed~up~(\alpha_1),Wait~(\alpha_2),Initiate~(\alpha_3),\\
Recover~(\alpha_4),Abandon~(\alpha_5), Maintain~(\alpha_6)\}$;

\item Cost function:
\begin{align}
J(s(k),a(k))=\begin{cases}c_1,~ ~s(k)=\zeta_1\\
	c_2,~ ~s(k)=\zeta_2\\
	c_3,~ ~s(k)=\zeta_3\\
        c_4,~ ~s(k)=\zeta_4\\
        c_5,~ ~s(k)=\zeta_5,
	\end{cases}
\end{align}
where $c_1<c_3<c_2<c_5<c_4$. 
Note that the state ``Cruise ($\zeta_1$)" indicates that the ego vehicle will ``Recover ($\alpha_4$)" to the reference signals (\emph{i.e.}, velocity and lane center) of the occupied lane after completing the lane change. 
Hence, the associated cost when the state $s=\zeta_1$ is the smallest.
``Return ($\zeta_5$)" implies that the ego vehicle must promptly return to its \emph{original lane} if the neighboring lane is deemed unsafe for lane change, which is different from ``Cruise ($\zeta_1$)".
Moreover, according to the Minimum Safety Spacing (MSS) analysis in the paper \cite{zhang2022integrated}, the preferred strategy for the ego vehicle to avoid traffic disruptions caused by emergency braking and vehicle acceleration is quick lane changing.
That is why the costs for ``Braking ($\zeta_2$)" and ``Acceleration ($\zeta_4$)" are greater than the cost for ``Quick lane change ($\zeta_3$)".
Emergency braking is only initiated when no feasible conditions for quick lane changing maneuvers exist, determined by the motion states of obstacle vehicles in the current lane and those in the adjacent lane.
In addition, engaging in ``Acceleration" may lead to a violation of traffic rules regarding speed limits.
While occasional deviations from the speed limit may be permissible for safety reasons, 
in typical circumstances, "Acceleration" should be the last resort for adhering to traffic rules.
This is why the associated costs ($c_5$) are often the highest.
\end{itemize}

The models of ego vehicle and surrounding vehicles are given as follows. 

\subsubsection{Ego vehicle model}

This paper mainly focuses on the high-level decision making
and the corresponding performance analysis. 
Therefore, a simplified model is used to represent the dynamics of the ego
vehicle as follows:
\begin{align}\label{exaegomodel}
x_{HV}(k+1)&=x_{HV}(k)+v_{HV}(k)T_h   \nn\\
v_{HV}(k+1)&=v_{HV}(k)+a_{HV}(k)T_h,
\end{align}
where $x_{HV}(k)$ and $x_{HV}(k+1)$ denote the longitudinal positions at time step $k$ and $k+1$, respectively, with a sampling time of $T_h$.
$v_{HV}(k)$ and $v_{HV}(k+1)$ represents the longitudinal speed at time step $k$ and $k+1$, respectively. 
$a_{HV}(k)$ is updated based on MDP states, which is given by 
\begin{align}\label{accupdate}
a_{HV}(k)=\begin{cases}0,~ ~s(k)=\zeta_1\\
	de,~ ~s(k)=\zeta_2\\
	0,~ ~s(k)=\zeta_3\\
        ac,~ ~s(k)=\zeta_4\\
        0,~ ~s(k)=\zeta_5,
	\end{cases}
\end{align}
where constants $de$ represents deceleration,
$ac$ denotes the acceleration on its original lane.

Additionally, 
when the \emph{Quick lane change} is not activated, 
the lateral position remains centered within the original lane.
However, when the \emph{Quick lane change ($\zeta_3$)} state is triggered, inspired by equation (6) presented in \cite{zhang2022integrated}, the lane-changing path is updated according to the following equation:
\begin{align}\label{yupdate}
y(t_{ref})= 10y^{d}_{qlc} \left(\frac{t_{ref}}{t^{d}_{qlc}}\right)^3-15y^{d}_{qlc} \left(\frac{t_{ref}}{t^{d}_{qlc}}\right)^4+6y^{d}_{qlc} \left(\frac{t_{ref}}{t^{d}_{qlc}}\right)^5,
\end{align}
where $t_{ref}$ is the time stamp since the lane changing maneuver is triggered with $t_{ref}\in [0,t^{d}_{qlc}]$.
Note that $t^{d}_{qlc}$ is the lane changing duration and $y^{d}_{qlc}$ is the lane center of the left lane. 

\subsubsection{Modelling surrounding vehicles}
In this paper, we assume that all other vehicles remain within their designated lanes without making lane changes. Furthermore, we acknowledge that all other vehicles have the flexibility to vary their speeds rather than maintaining a constant velocity. 
Therefore, we employ the following straightforward double-integrator model to describe the behavior of the surrounding vehicles:
\begin{align}\label{exasurroundingmodel}
x_{j}(k+1)&=x_{j}(k)+v_{j}(k)\Delta t+\frac{1}{2}a_j(k)(\Delta t)^2\nn\\
y_{j}(k+1)&=y_{j}(k)\nn\\
v_{j}(k+1)&=v_{j}(k)+a_{j}(k)\Delta t,
\end{align}
where $x_j(k) ~(y_j(k)), v_j(k)$ and $a_j(k)$ represent longitudinal position (lateral position) in the global coordinate, velocity and acceleration of surrounding Vehicle $j$ at time step $k$, respectively. 
The sampling time, denoted as $\Delta t$, is equivalent to $T_h$ for high level decision making and $T_l$ for low level path planning.
This is because we consider using the same double-integrator model of surrounding vehicles no matter for high-level decision making or for low-level path planning in this application.

\subsubsection{Safety constraints}

\begin{itemize}
    \item if ego vehicle is on the original lane, $x_{Or}(k)-x_{HV}(k)\geq d_{safe}$, where $Or$ is the front vehicle (see the following Fig. \ref{ex2initialposition});

    \item if $|x_{HV}(k)-x_j(k)|\leq d_{safe}$, $s(k+1)\neq \zeta_3 (Quick~ lane ~change)$, where $j\in\{Ob, Og\}$ is the Vehicle on the adjacent lane (see the following Fig. \ref{ex2initialposition}).
\end{itemize}

Next, based on the formulation mentioned above, we will explore a case  to showcase the capability of the proposed framework in managing diverse behaviors within intricate and ever-changing environments.

\subsection{Case Study}

To run simulations, the following parameters are given:
\begin{enumerate}

  \item MDP states and actions are numerical as
\begin{itemize}
 \item $\zeta_1=1, \zeta_2=2,\zeta_3=3,\zeta_4=4,\zeta_5=5$
 \item $\alpha_1=6, \alpha_2=7,\alpha_3=8,\alpha_4=9,\alpha_5=10,\alpha_6=11$;
\end{itemize}
  \item  Prediction horizon and sampling time
\begin{itemize}
    \item $N_h=4, T_h=0.4(s)$ for the high-level decision making;
     \item $N_l=3, T_l=0.1(s)$ for the low-level path planning;
\end{itemize}
That is, in the high level, the decisions will be updated
every 4 steps. 
\item Other parameters: $c_1=0,c_2=3,c_3=2,c_4=10,c_5=5$, $d_{safe}=15 (m), de=-4 (m/s^2),ac=4(m/s^2)$.
\end{enumerate}
\begin{remark}
In this paper, our primary focus lies in high-level decision-making, and we make the assumption that at the low level, the controller has been meticulously designed and is capable of perfectly executing the commands originated from the high-level decisions. 
As such, to facilitate a meaningful comparison with the rule-based method presented in \cite{zhang2022integrated}, we consider using the rules given in \cite{zhang2022integrated} to calculate the terminal cost for Algorithm 1. 
\end{remark}

\begin{figure}[htb!]
	\centering
	\includegraphics[width=7cm]{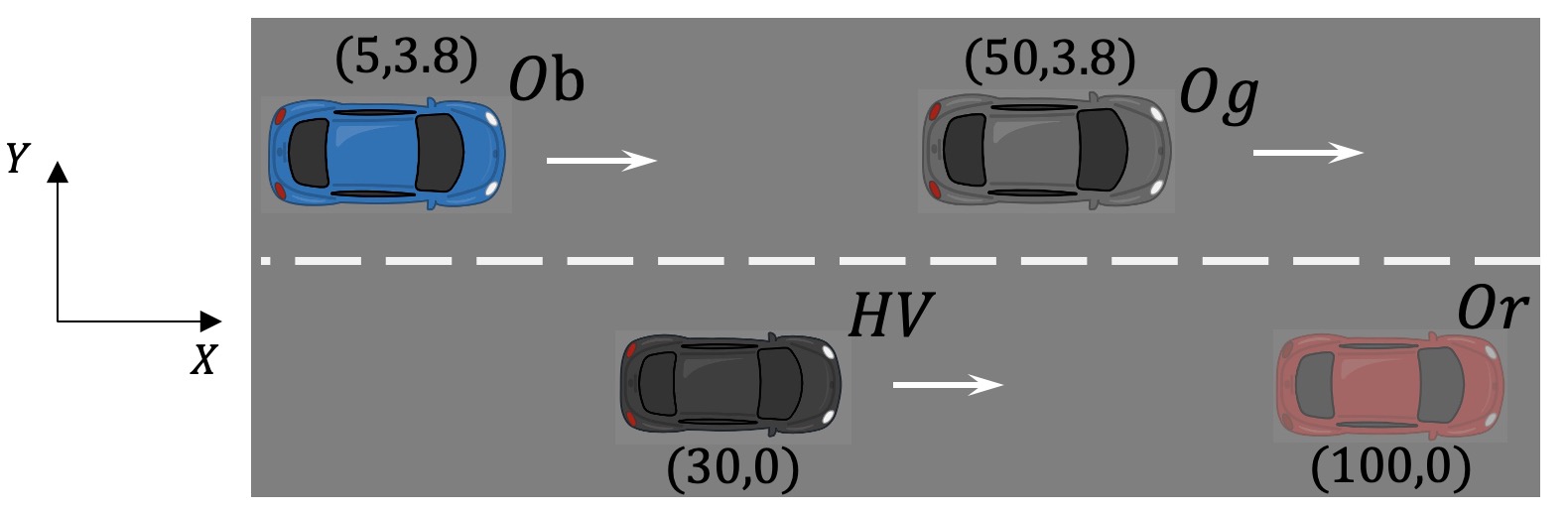}
	\caption{Initial positions of the ego vehicle (\emph{i.e.}, HV) and its surrounding vehicles. The unit of the numbers in the figure is meter.}
	\label{ex2initialposition}
\end{figure}

To demonstrate that our proposed MPC-based HMDP control framework is significantly safer in complex and dynamic environments, we now present a  scenario in which the rear blue vehicle (denoted as `Ob' in Fig. \ref{ex2initialposition}) approaches with varying speeds over time.

\emph{\textbf{Scenario: Rear vehicle is moving with varying speeds}}

\begin{figure}[htb!]
	\centering
	\includegraphics[width=6cm]{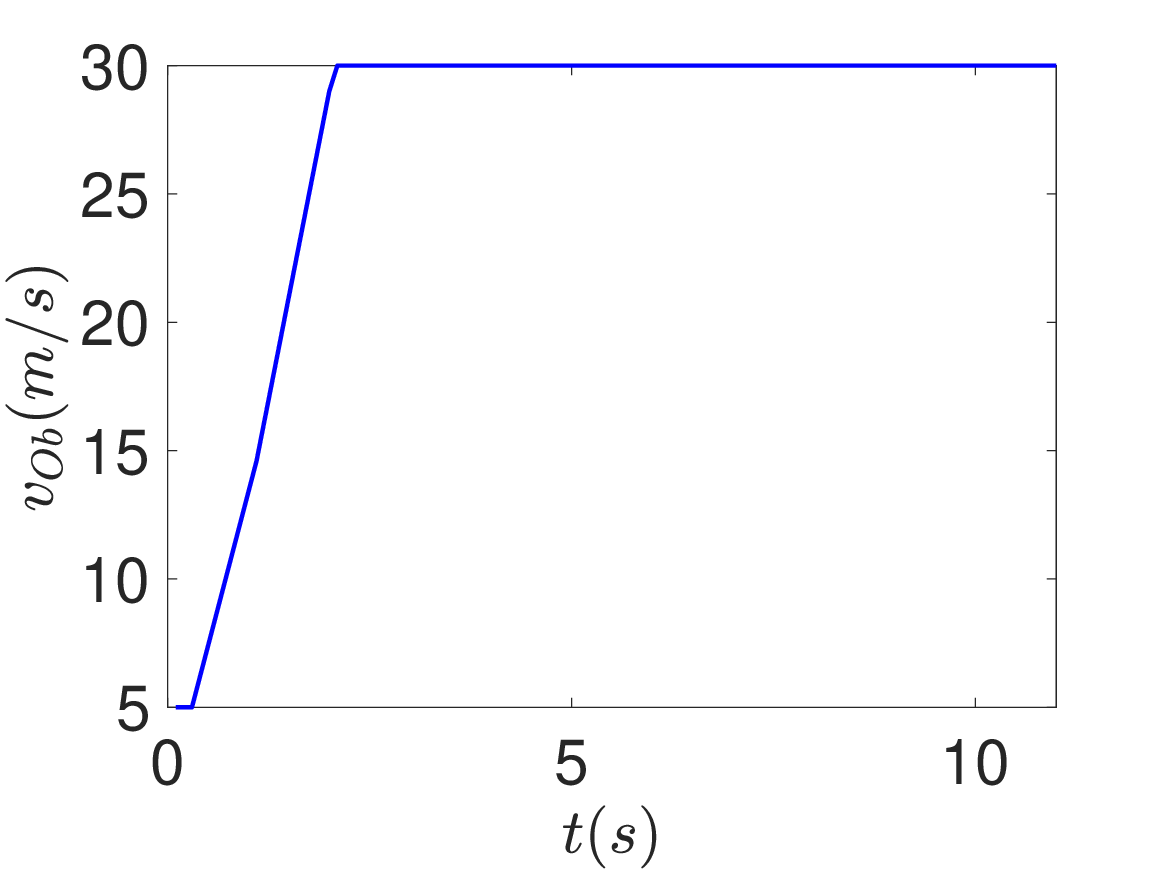}
	\caption{The speed of rear blue vehicle.}
	\label{ex2speedrearvehicle}
\end{figure}

\begin{figure}[htb!]
	\centering
	\includegraphics[width=7cm]{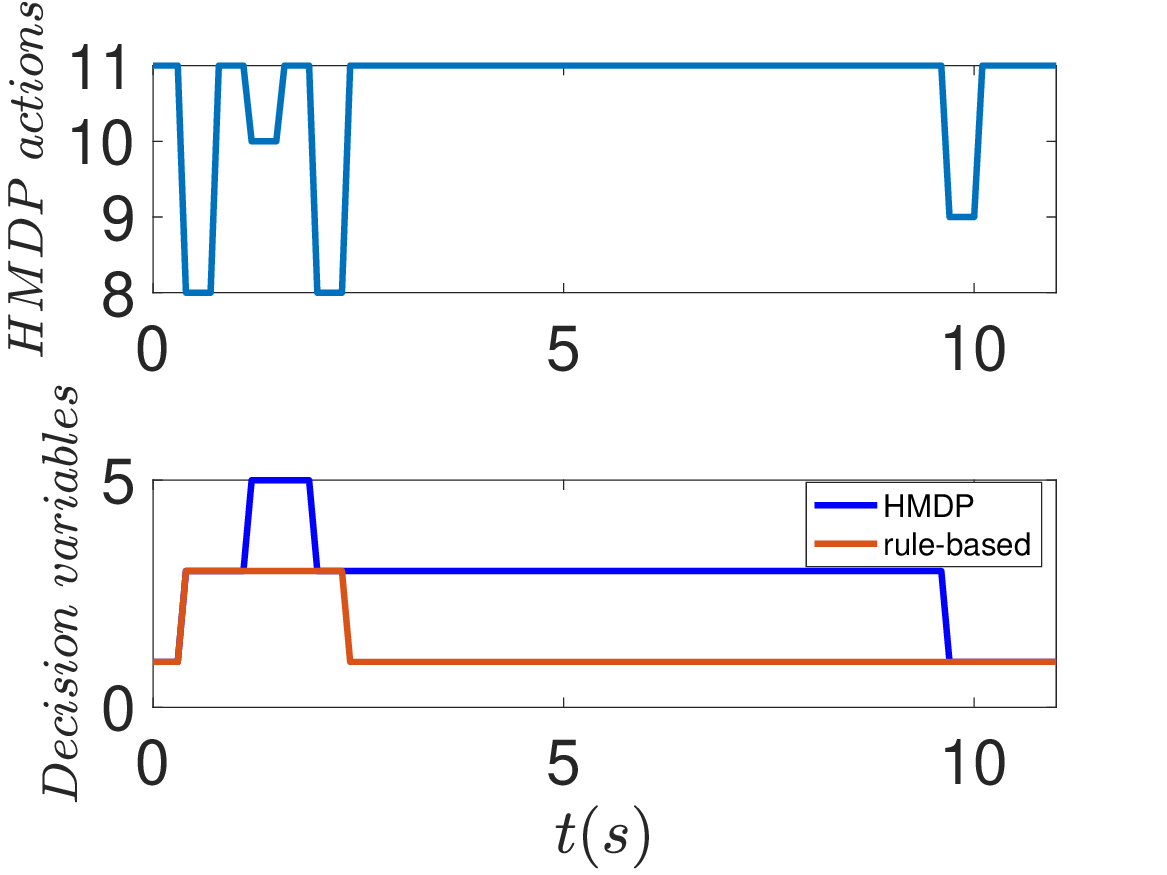}
	\caption{HMDP-based state transitions.}
	\label{ex2MDPactions}
\end{figure}

\begin{figure}[htb!]
	\centering
	\includegraphics[width=7cm]{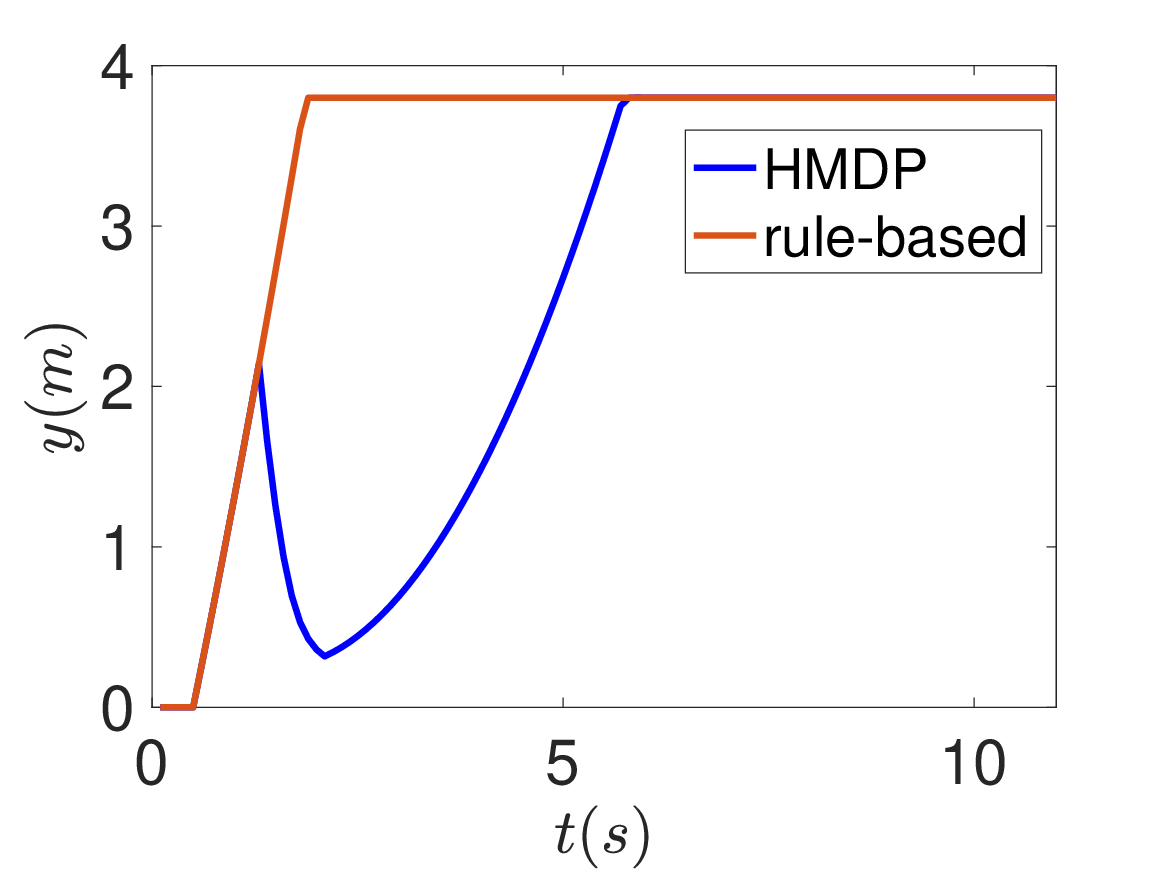}
	\caption{The lateral position of ego vehicle under two methods.}
	\label{ex2MDPy}
\end{figure}

\begin{figure}[htb!]
	\centering
	\includegraphics[width=8cm]{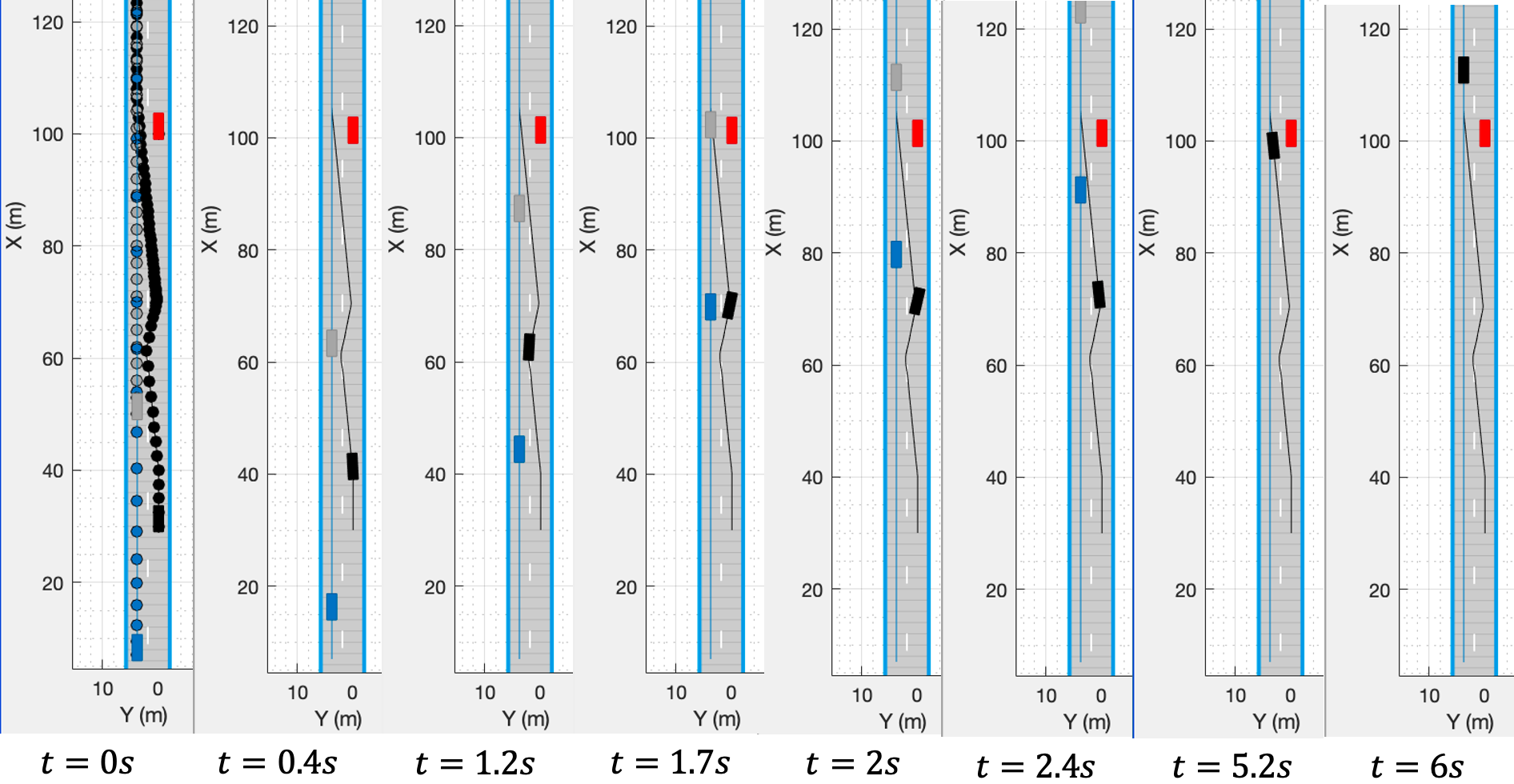}
	\caption{A series of screenshots for lane changing process under HMDP-based control framework.}
	\label{ex2MDPscreenshots}
\end{figure}

\begin{figure}[htb!]
	\centering
	\includegraphics[width=7cm]{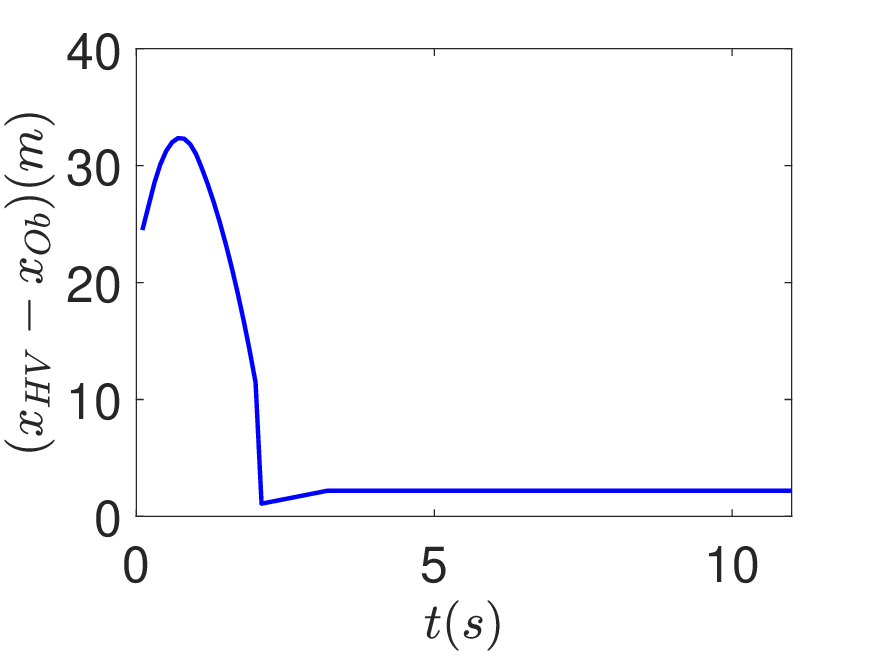}
	\caption{Gap between the ego vehicle and the rear vehicle under rule-based method.}
	\label{gap}
\end{figure}

\begin{figure}[htb!]
	\centering
	\includegraphics[width=7cm]{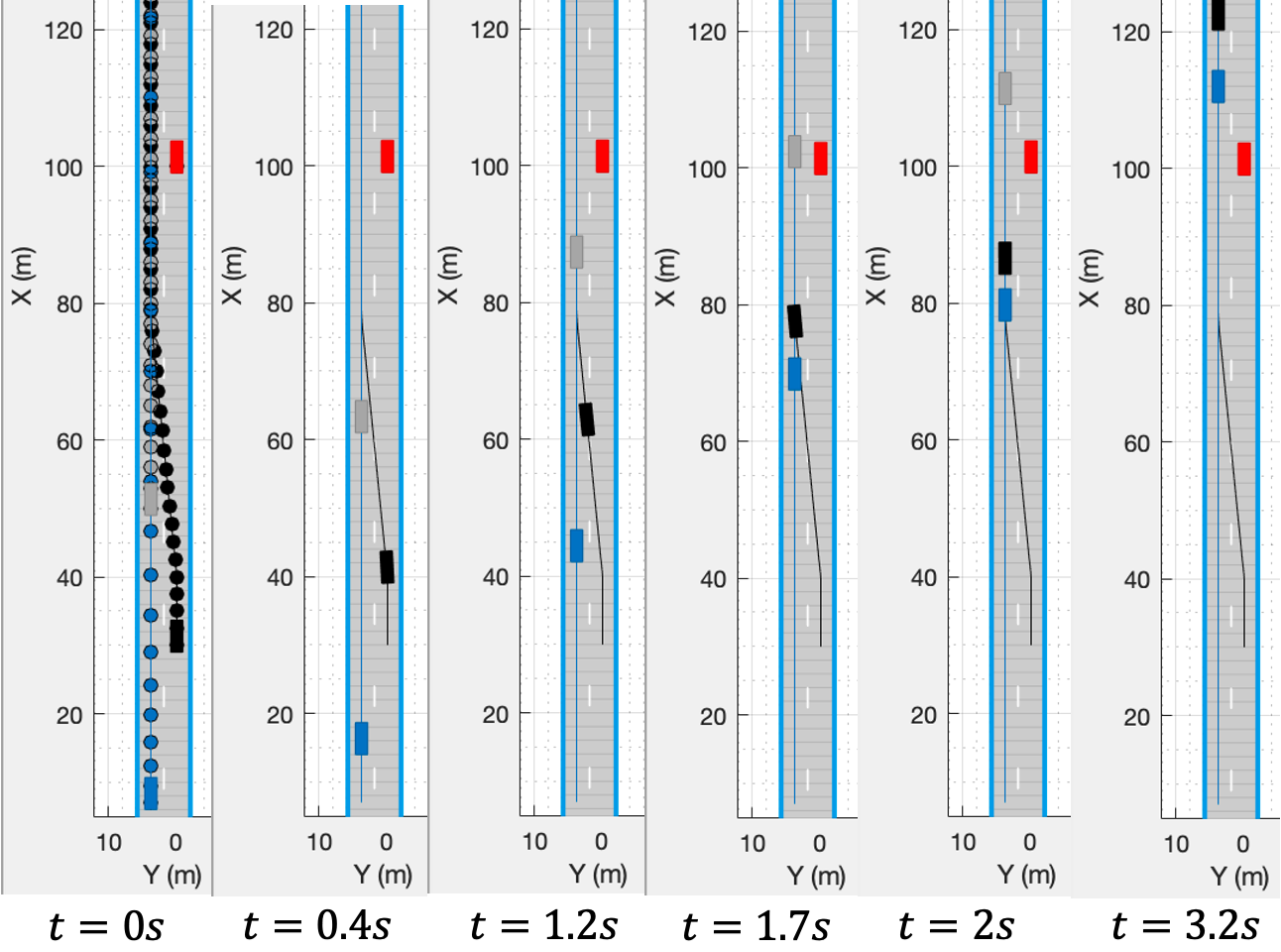}
	\caption{A series of screenshots for lane changing process under rule-based method.}
	\label{ex2rulescreenshots}
\end{figure}

In this scenario, the initial positions of these vehicles are shown in 
Fig. \ref{ex2initialposition}.
The initial velocity of the ego vehicle is 
$v_{HV}=25 (m/s)$ and other vehicles with constant speeds are respectively given as follows:
$v_{Or}=0 (m/s), v_{Og}=30 (m/s)$.
The speed of rear blue vehicle is time varying as given in Fig. \ref{ex2speedrearvehicle}.
We set the overall simulation time to 11(s). The simulation results are shown in Fig. \ref{ex2MDPactions}-Fig. \ref{ex2rulescreenshots}.

From Fig.~\ref{ex2initialposition} and Fig.~\ref{ex2speedrearvehicle} we know that at the beginning, the rear blue vehicle (\emph{i.e.}, Ob) is moving slowly from far distance. 
So Fig.~\ref{ex2MDPactions} shows that high-level framework makes the decision for ego vehicle to initiate the lane change at time $t=0.4 s$ (\emph{i.e.,} $a(t=0.4 s)=8 (Initiate), s(t=0.4 s)=3 (Quick ~lane ~change)$).
However, Vehicle $Ob$ speeds up suddenly during the lane changing process (see Fig. \ref{ex2speedrearvehicle}).
To ensure passenger safety, the ego vehicle appears to make a prudent decision to promptly return to its original lane, as inferred from the information presented in Figs. \ref{ex2MDPactions}-\ref{ex2MDPy} (\emph{i.e.,} $a(t=1.2 s)=10 (Abandon), s(t=1.2 s)=5 (Return)$)). 
This action allows ample space for the trailing vehicle to pass safely.
After $0.8 s$, the rear vehicle passes and high-level framework makes the new decision for ego vehicle to initiate the lane change again (\emph{i.e.,} $a(t=2 s)=8 (Initiate), s(t=2 s)=3 (Quick~ lane~ change)$. 

Contrarily, based on the data depicted in Fig. \ref{ex2MDPactions}, Fig. \ref{ex2MDPy}, Fig.\ref{gap} and Fig. \ref{ex2rulescreenshots}, it appears that the ego vehicle persists in changing lanes using a pre-defined rule at the time $t=1.2 s$, even as the rear vehicle accelerates.
As a result, starting from the time $t=1.7 s$, the gap between the ego vehicle and the rear vehicle appears to be diminishing (see 
Fig.\ref{gap}), which could potentially cause unease among the passengers.
In the predefined rule, lane changes are restricted by a specific duration.
Once a lane change is initiated, it cannot be reversed. 
This approach may prove ineffective when dealing with dynamic environments.
The proposed HMDP-based control framework in this paper possess a degree of adaptability as they can update models or optimize objectives by exploring and exploiting to accommodate changing environments.
Videos demonstration of this case based on HMDP and rule methods are available at \emph{https://youtu.be/GNZisb9TYPU} and \emph{https://youtu.be/MQJV0PRrIDI}, respectively. 

\begin{remark}
In this case study, we use MPC-based path planner at the low level with quadratic cost function for reference tracking.
Moreover, the cost function may vary according to the high level decision making (\emph{i.e.}, MDP state $\zeta$).
In addition, we utilize elliptic safety constraints with respect to the surrounding vehicles (\emph{e.g.,} leading vehicle ($Or$) in the current lane, rear vehicle ($Ob$) and front vehicle ($Og$) in the adjacent lane, see Fig. \ref{ex2initialposition}).
\end{remark}

\section{Conclusions}
\label{conclusions}

This paper has presented a unified hierarchical control framework by integrating MDP and MPC for autonomous decision-making systems.
After carefully formulating the top level of decision making as an HMDP control problem, an MPC-based scheme is designed to optimise the decision for autonomous systems while ensuring safety.
By integrating low-level control systems into high-level decision-making processes, the framework has ability to generate correct decisions in complex environments. 
This means that through the combination of HMDP and MPC, the proposed solution guarantees both safety and optimality, as showcased through simulations and applications. 
Furthermore, the framework offers a promising approach for designing autonomous decision-making systems capable of handling dynamic and uncertain environments.
To establish recursive feasibility and stability of the novel framework, the terminal cost covering cost-to-go is calculated under a mild assumption and it is involved in online solving the HMDP optimisation problem.
Finally, an autonomous lane-changing scenario is explored to showcase the impressive capabilities of the proposed MPC-based HMDP control framework in managing a wide range of behaviors within dynamic and intricate environments. Our future work is to extend this HMDP modelling and MPC based analysis and design framework to a wider range of decision making problems arising in autonomous systems operating under a dynamic and uncertain environment to ensure safety and optimality.

\bibliographystyle{plain}
\bibliography{fmref}

\end{document}